\documentclass[aps,pre,twocolumn,superscriptaddress,groupedaddress,showpacs,amsmath,amssymb,nofootinbib,scrartcl]{revtex4-1}
\usepackage{graphicx}
\usepackage{graphics}
\usepackage{amsmath}
\usepackage{subfigure}
\usepackage{color}
\usepackage{url}
\usepackage[colorlinks=true,linkcolor=red,citecolor=blue,filecolor=black,urlcolor=black]{hyperref}
\begin{document}

\title{Phononic Topological States in 1D Quasicrystals}
\author{J. R. M. Silva}
\affiliation{Departamento de F\'{\i}sica Teórica e Experimental, Universidade Federal do Rio Grande do Norte,59078-900, Natal--RN Brazil}
\author{M. S. Vasconcelos}
\email[Corresponding author:\ ]{mvasconcelos@ect.ufrn.br}
\altaffiliation[Permanent address: ]
{Escola de Ci\^encias e Tecnologia, Universidade Federal do Rio Grande do Norte, 59078-900, Natal--RN, Brazil}
\affiliation{Department of Physics and Astronomy, Western University, London--Ontario, N6A 3K7, Canada}
\date{Received: date / Revised version: date}
\author{D. H. A. L. Anselmo}
\affiliation{Departamento de F\'{\i}sica Teórica e Experimental, Universidade Federal do Rio Grande do Norte,59078-900, Natal--RN Brazil}
\date{\today}
\author{V. D. Mello}
\affiliation{Departamento de F\'{\i}sica, Universidade do Estado do Rio Grande do Norte, Mossor\'{o}--RN, 59625-620, Brazil}

\begin{abstract}
	We theoretically analyze the spectrum of phonons of a one-dimensional quasiperiodic lattice. We simulate the quasicrystal from the classic system of spring-bound atoms with a force constant modulated by the Aubry-André model, so that its value is slightly different in each site of the lattice. From the equations of motion, we obtained the equivalent phonon spectrum of the Hofstadter butterfly, characterizing a multifractal. In this spectrum, we obtained the extended, critical and localized regimes, and we observed that the multifractal characteristic is sensitive to the number of atoms and the $\lambda$ parameter of our model. We also verified the presence of border states for phonons, where some modes in the system boundaries present vibrations. Through  the measurement of localization of the individual displacements in each site, we verify the presence of a phase transition through the Inverse Participation Rate (IPR) for $\lambda= 1.0 $, where the system changes from extended to localized.

\end{abstract}



\maketitle

\section{Introduction}

The discovery of a new class of materials by Shechtman \textit{et al.} in 1984 \cite{PhysRevLett.53.1951}, when they were studying the diffraction figures for an alloy of Aluminum and Manganese ( that give him the Nobel Prize in Chemistry of 2011 \cite{NOBEL-qcrystal}), had started up a new and very rich research  area.  At the first view, this system was defined as an intermediate structure between crystalline and amorphous solids, but today is well recognized that quasicrystals, is interpreted as a natural extension of the notion of a crystal to structures with quasiperiodic (QP), instead of periodic, arrangements of atoms \cite{PhysRevLett.53.2477,Enrique2014}. A more recent updated definition of quasicrystals with dimensionality {\it n} ({\it n} = 1, 2 or 3) is that they can also be defined as a projection of a periodic structure in a higher dimensional space $mD$, where $m >n$ \cite{Vardeny}.
The diffraction figure found by Schectman and collaborators has a long-range order but has no translational periodicity as the crystals, but rather the self-similarity property by scaling \cite{PhysRevLett.53.1951}. In the icosahedral and decagonal quasicrystals, the self-similarity is related to the \textit{Golden Ratio} $((1 + \sqrt{5})/2)$, so that the atoms are separated by distances that represent the Fibonacci sequence. These new materials have great potential for applications, as some researches show that they are rigid and brittle with unique transport characteristics  \cite{RECHTSMAN2008} and have very low surface energies that make them good thermal insulators with photonic and thermoelectric properties \cite{SORDELET1997,HIPPERT1994,DUBOIS2000}. Its spectrum features a fractal structure and aspects of electronic localization \cite{MVASCONCELOS2007,MVASCONCELOS2011} and optical \cite{MVASCONCELOS1998, MVASCONCELOS1999, MVASCONCELOS2007}. The growing studies on quasicrystalline materials made it possible to obtain systems with very thin layers arranged in a quasiperiodic sequence \cite{PhysRevLett.90.055501}or wires with quantum wells of width around 7 nm\cite{PhysRevLett.112.146404}.
Levine \textit{et al.}  works \cite{LEVINE19841D}, with the synthesis of a quasicrystal defined by the Fibonacci sequence inspired Merlin \textit{et al.} to create, in the laboratory, the first one-dimensional quasicrystal \cite{MERLIN19851D}. Since that, the way of studying one-dimensional quasicrystals  structures has become standard. In this procedure we define two distinct building blocks, each of which carries the necessary physical information, and then they are arranged according to a particular sequence. For example, they can be described in terms of a series of generations that obey the relation of particular recursion\cite{MVASCONCELOS1999}.The researches have shown fractal properties in their spectra and the existence of a non-trivial phase transition, such as the metal-insulating phase \cite{DalNegro2003, Kyek2000} only by adjusting some parameters of the generation sequence. Recently, it was reported that it is possible to have a topological phase in photonic quasicrystals (in 2D) without any magnetic field applied, but instead introducing an artificial gauge field via dynamic modulation \cite{PhysRevX.6.011016}. The idea that photonic crystals could exhibit an analog like the quantum Hall edge states was initially proposed by Haldane and Raghu\cite{PhysRevLett.100.013904} in tridimensional photonic crystals.

On the other hand, phononic crystals are intensively studied as means to manipulate sound or elastic waves (for review see \cite{ISI:000337909500003}) in the same way like in photonic crystals.  Following the same idea of Haldane and Ragh, researches have given attention to the search of the existence of topologically protected edge states \cite{PhysRevLett.115.104302,ISI:000387480100001} in those systems, which could be beneficial for many practical  applications\cite{PhysRevB.96.134307,Ni_2015,ISI:000364919700001}. The topological effects in the band structure in one-dimensional phononic crystal can be characterized through topological invariants such as Berry phases \cite{RevModPhys.82.1959,PhysRevLett.62.2747} and Zak phases\cite{ISI:000327944600018,ISI:000378949500001,ISI:000431113100004}. Recently, it was shown that these edge modes exist in the band gap of 1D phononic crystals composed of two different phononic crystals, with distinct topological properties\cite{ISI:000379070900001}. It also has been founded in other types of phononic cystals\cite{Susstrunk47}. In this work, we investigate a system where it is possible to have these edge modes in 1D, i.e., we study the edge modes in 1D phononic quasicrystals.

Indeed,  1D quasicrystals were studied through the Harper model \cite{KRAUS2012,HARPERDISORD1958}. Many works on one-dimensional quasicrystals showed that the localization property of the Harper model \cite{HARPERDISORD1958} could be found in a quasicrystal through the Hamiltonian of Aubry-André, considering the potential incommensurable with the lattice parameter \cite{BIDDLE2011-IPR,GANESHAN2013-EZERO,ROSAS1998}.  This model proved to present itself as a topological insulator which exhibits border states and non-trivial phases, experimentally verified in the works of Kraus \textit{et al.} \cite{KRAUS2012}, which used waveguides to obtain the frequency spectrum in a quasicrystal, indicating the existence of a photonic gap \cite{ZOOROB2000, ZHANG2001}. The vast majority of published papers use superlattices, showing a fragmented energy spectrum of the famous Hofstadter butterfly \cite{HOFSTADTER1976} at the electronic level, as well as for the optical case \cite{LANG2012}.
\par Theoretical models for predicting the properties of quasiperiodic systems have been of considerable interest to the scientific community, resulting in many theoretical and experimental studies \cite{BABOUX2017,WANG2018,JeevanH.S.2004,KRAUS2012,MAN2003}. However, some properties, like edges modes and topological states, in one-dimensional quasicrystals remain unexplored. The localization of phonons in one-dimensional lattices has already been studied for the Frankel-Kontorova model \cite{Kontorova1938, FK-AUBRY-DISCRETE1983} and for quasiperiodic systems by the transfer matrix formalism \cite{BURKOV1996, SALAZAR2003}. Some works compare the frequency spectrum with the energy bands obtained for a quasicrystal defined by the transfer matrix formalism \cite{KOHMOTO1986,YOU1990}, in which the results retain fractality properties. However, these studies do not consider the effects of the initial phase $\phi$, as in the works of  Kraus \textit{et al.} \cite{KRAUS2012}.
Therefore, in this work, we have studied the frequency spectrum, and localization of phonons in a quasiperiodic lattice through a modulation of the Aubry-André  \cite{AUBRY1980} model in order to characterize these edge states as phononic topological states, and also we studied the analogous to metal-insulator phase transition for this system.

This paper is organized in the following way. In section 2, we present the theoretical model for the 1D Quasicrystals studied here. In section 3 we show our numerical results and discussion. First, we present a profile spectra similar to the Hofstadter butterfly \cite{HOFSTADTER1976}. After, we present the topological states of phonons, considering an equivalent model studied by Kraus \textit{et al.} \cite{KRAUS2012}, showing the presence of border states for phonons. At last, we study the phase transition through the Inverse Participation Rate (IPR), where the system changes from extended to localized. Finally, in section 4, we present the conclusions of this paper.

\section{Theoretical Model}

The simple model for the phonon system can be defined through the motion equation that represents the atoms as a spring-bound system with force constant $K _{n}$.
\begin{eqnarray}
 \omega^2 u_{n} =-K_{n + 1} u_{n + 1}-K_{N} u_{n-1} + V_{n} u_n
\end{eqnarray}
Here, $\omega$ matches the vibration frequency, $u_{n} $ are the individual displacementes around the equilibrium position and $V_{n} = K_{n + 1} + K_{n-1}$.

Considering a system of $N$ atoms the motion equation has the following matrix form:

\begin{eqnarray}\label{eq:matrix-fonon}
\begin{pmatrix}
V_{0} & -K_{n} & 0 & 0\\
-K_{n} & V_{1} & -K_{n+1} & 0\\
0  &\ddots&\ddots &\colon \\
0 	& 0 & -K_{N+1} & V_{N}
\end{pmatrix}
\begin{pmatrix}
u_{0}\\
u_{1} \\
\colon\\
u_{N}
\end{pmatrix}=\omega^2\begin{pmatrix}
u_{0}\\
u_{1} \\
\colon\\
u_{N}
\end{pmatrix}
\end{eqnarray}
The $u_{n}$ represent the position of successive atoms in the chain at a given site $n$, vibrating around the equilibrium position with frequency $\omega$.

\begin{figure}[ht!]
\centering
\includegraphics[width=1.0\linewidth]{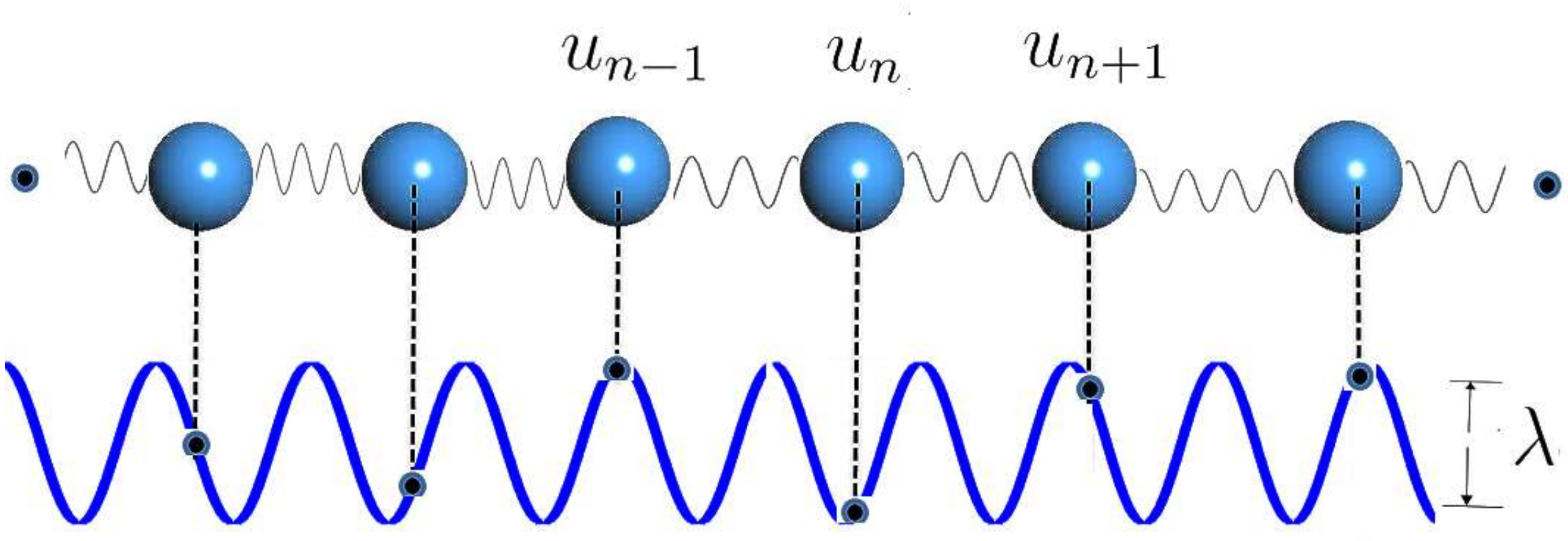}
\caption{Quasiperiodic lattice with the interaction of the potential of Aubry-André. Each atom is subject to a force constant $K_n$, incommensurate with the lattice parameter.}
\label{fig:fk-system}
\end{figure}

The force constant $K_n$ obeys a quasiperiodic modulation, undergoing small changes dependent on each site, as shown in the Figure \ref{fig:fk-system}.
Each atom has a mass $m_{n} = 1.0$, connected by springs with force constant $K_{n}$, given by:
\begin{equation}\label{eq:string}
K_n = C(1+\lambda \cos(2\pi b n +\varphi)); \qquad\ \  \beta=\frac{1+\sqrt{5}}{2}
\end{equation}
The sine term of $K_{n}$ comes from the interaction with the external potential of amplitude $\lambda$ on the force constant $C$. The $\varphi$ variable corresponds to the initial phase when $n = 0$. This type of system presents different results, depending on whether the inverse of the period is rational or irrational \cite{FKVANOSSI2000}.

From the eigenvalues of the motion equation \eqref{eq:matrix-fonon} we can obtain the frequency spectrum, while the eigenvectors give us the individual displacements for each site. The nature of these displacements $u_{n}$, similar to what occurs with the wave-function, can show itself as distributed over all sites or located in just a few ones \cite{BURKOV1996}. If this location is concentrated at the edge of the system, these correspond to the topological states of the border.
\par The location of the displacements can be obtained by the inverse of the participation rate (IPR) \cite{BIDDLE2011-IPR}. The IPR of the eigenvector $k$ can be obtained by the following relation:
\begin{eqnarray}
\text{IPR$_{(k)}$} = \displaystyle\frac{\sum_{l}|u_{k,l}|^{4}}{\left(\sum_{l}|u_{k,l}|^{2}\right)^{2}},
\end{eqnarray}
where the $l$ index represents the sum over all sites on the lattice. The IPR indicates the inverse of the number of occupied sites $L$, so when the oscillations are equally distributed, the IPR $\approx 1/L$, whereas on the opposite situation of extreme localization, we have only one site vibrating with the respective frequency, which results in IPR $\approx 1$.

The numerical results were obtained from the diagonalization equation \eqref{eq:matrix-fonon}, with unitary values for both the mass $(m_{n} = 1.0)$ and the force constant amplitude $(C = 1.0)$. In this way, we can study the influence of modulation on the force constant $K_n$, on the frequency spectrum of the phonons in a quasicrystal.
\par From the diagonalization of the phonon system matrix \eqref{eq:matrix-fonon}, we found the frequency spectrum, in order to analyze the propagation of phonons in this quasiperiodic media. In the works of You, J. Q \textit{et al}, this model was studied for the case in which the masses of successive atoms obey a Fibonacci sequence, using the formalism of the transfer matrix \cite{YOU1990}. They showed that the spectrum is truncated in a fractal to a larger amount of atoms, also evidenced in the works of Kohomoto \textit{et al} \cite{KOHMOTO1986} and in the works of F. Salazar \textit{et al} \cite{SALAZAR2003}, which proposed a modulation in the equation of motion. In our model, we used the computational package of \textit{Gnu Scientific Library} (GSL)\cite{GSL2009}, implemented in C++ routines to find the properties of the phonon spectrum in quasiperiodic media from a sine-type modulation in the force constant between neighboring atoms, with a varying $\varphi$ phase.

\section{Results and Discussion}

The phonon spectrum in the motion equation with force constant modeled in \eqref{eq:string} presents a profile similar to the Hofstadter butterfly, when plotted as a function of the inverse of the period $b$ \cite{HOFSTADTER1976}.

\begin{figure}[ht!]
\centering
\includegraphics[width=1.0\linewidth]{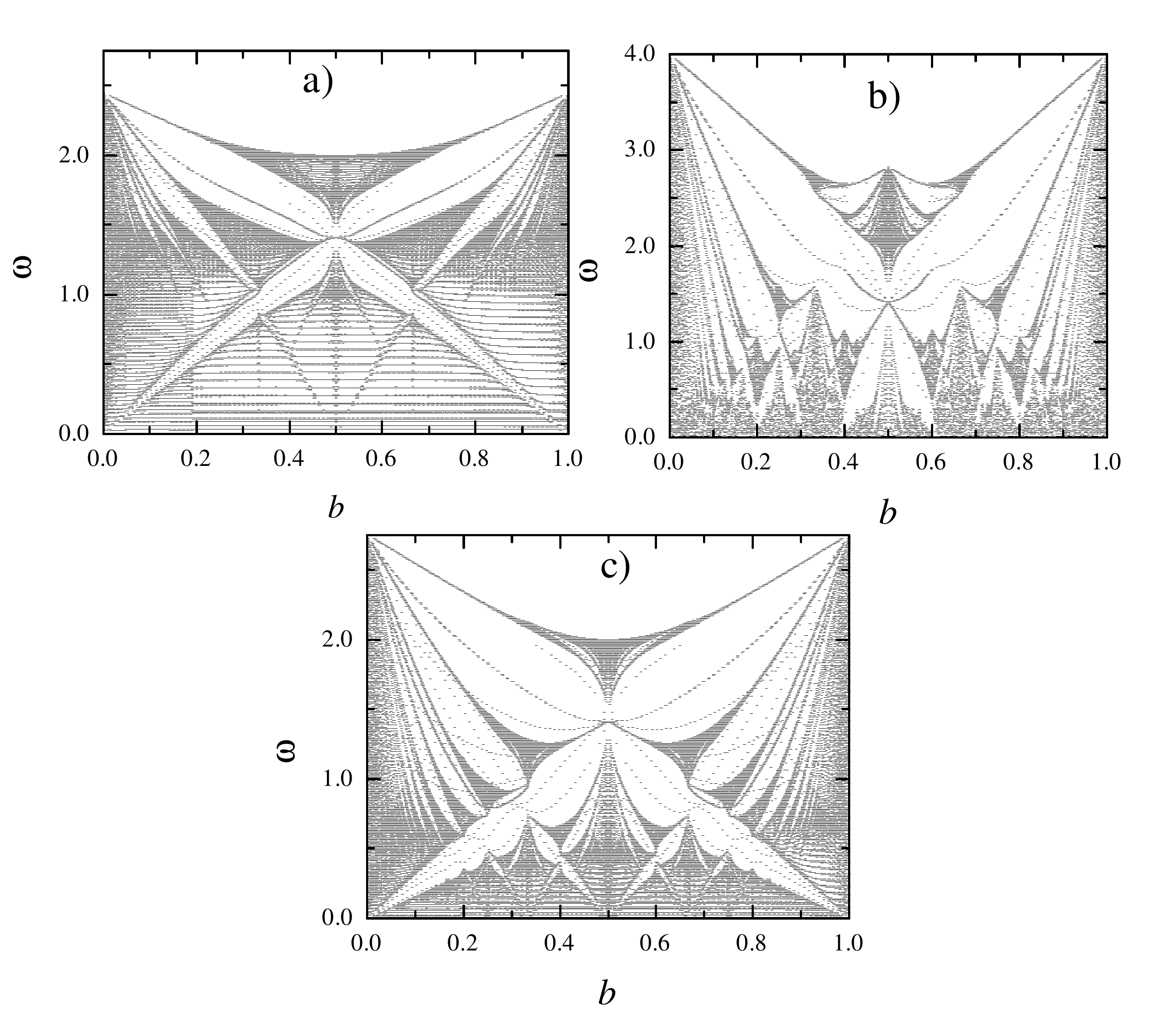}
\caption{Hofstadter butterfly for the frequency spectrum in three different  $\lambda$ values for the quasicrystalline potential with 100 atoms ($N = $100) and $\varphi=0$. We can notice that in $\lambda=1.0$ the spectrum presents a considerable amount of bands separated by increasingly narrow gaps characterizing a multifractal behavior.}
\label{fig:ph-but-lbda}
\end{figure}
The equation \eqref{eq:string} depicts a spectrum of the quasicrystal, shown when $b = \beta $. In Figure \ref{fig:ph-but-lbda} we have the phonon spectrum for three different values of the potential amplitude $\lambda$ as a function of the inverse of the period $b$. In $\lambda=0.5$  the spectrum presents bands very close while for $\lambda=1.0$  we have the parameter which represents a critical system \cite{Wilkinson1984}. For this value of $\lambda$,  the allowed frequencies are defined by several bands composed of increasingly narrow gaps located between the four larger gaps, characterizing a multifractal spectrum, as we will see further in Fig. 3. As $\lambda$  grows,  there is a like deformation between the gaps and the spectrum loses this characteristic, and  we can see the deformation of the larger gaps for $\lambda = 3.0$ due to intense variations in the force constant (Fig. \ref{fig:ph-but-lbda} on top at right ).
\par
\begin{figure}[ht!]
\centering
\includegraphics[width=1.0\linewidth]{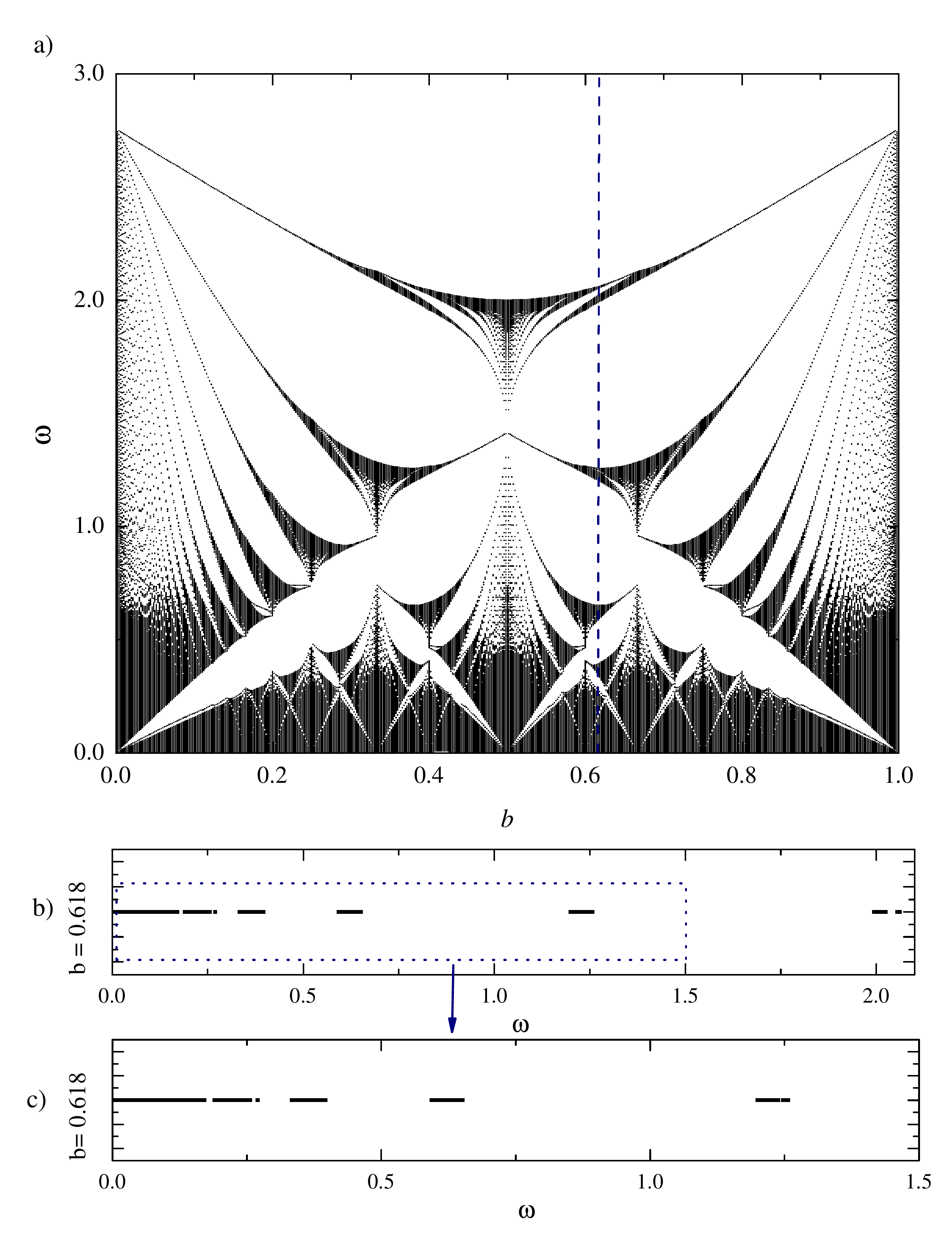}
\caption{Hofstadter butterfly for the frequency spectrum in a quasicrystal as a function of parameter $b$. We use $N = 400$, $\varphi = \pi/2$ and $\lambda = 1.0$. We highlight (red in online color version) the expansion in the region where $b \approx 0.618$ to evince the frequency replication pattern, where the spectrum repeats in a self-similar fashion.}
\label{fig:ph-but-lbda-one}
\end{figure}

 On the other hand, the states that cross larger gaps are sensitive to the number of atoms in the lattice \cite{KRAUS2012}. Inspired in previous works, we have used a number of $N = $100 sites, and we obtained a spectrum with certain bands circumventing the larger gaps. In Figure \ref{fig:ph-but-lbda-one} we show the frequency spectrum as a function of the inverse of the period $b$, for $400$ sites and setting the phase $\varphi =  \pi/2$ to obtain higher definition in the frequency spectrum of phonons.

We can see that the figure is similar to the Hofstadter butterfly obtained for the electronic case of the Hamiltonian of Aubry-André \cite{AUBRY1980,MADSEN2013}. The variations of $b$ present a characteristic of self-similarity for the frequencies, maintaining the structure composed by the larger gaps and some crossed modes. In the highlighted region we see that the three major gaps are replicated, and the frequency spectrum follows the same pattern. The limit for this replication is ruled by the precision of step, $b$, where we consider an increment of $10^{-3}$ for a fixed phase $\varphi =  \pi/2$ in the equation \eqref{eq:string}.
\par
\begin{figure}[ht!]
\centering
\includegraphics[width=1.0\linewidth]{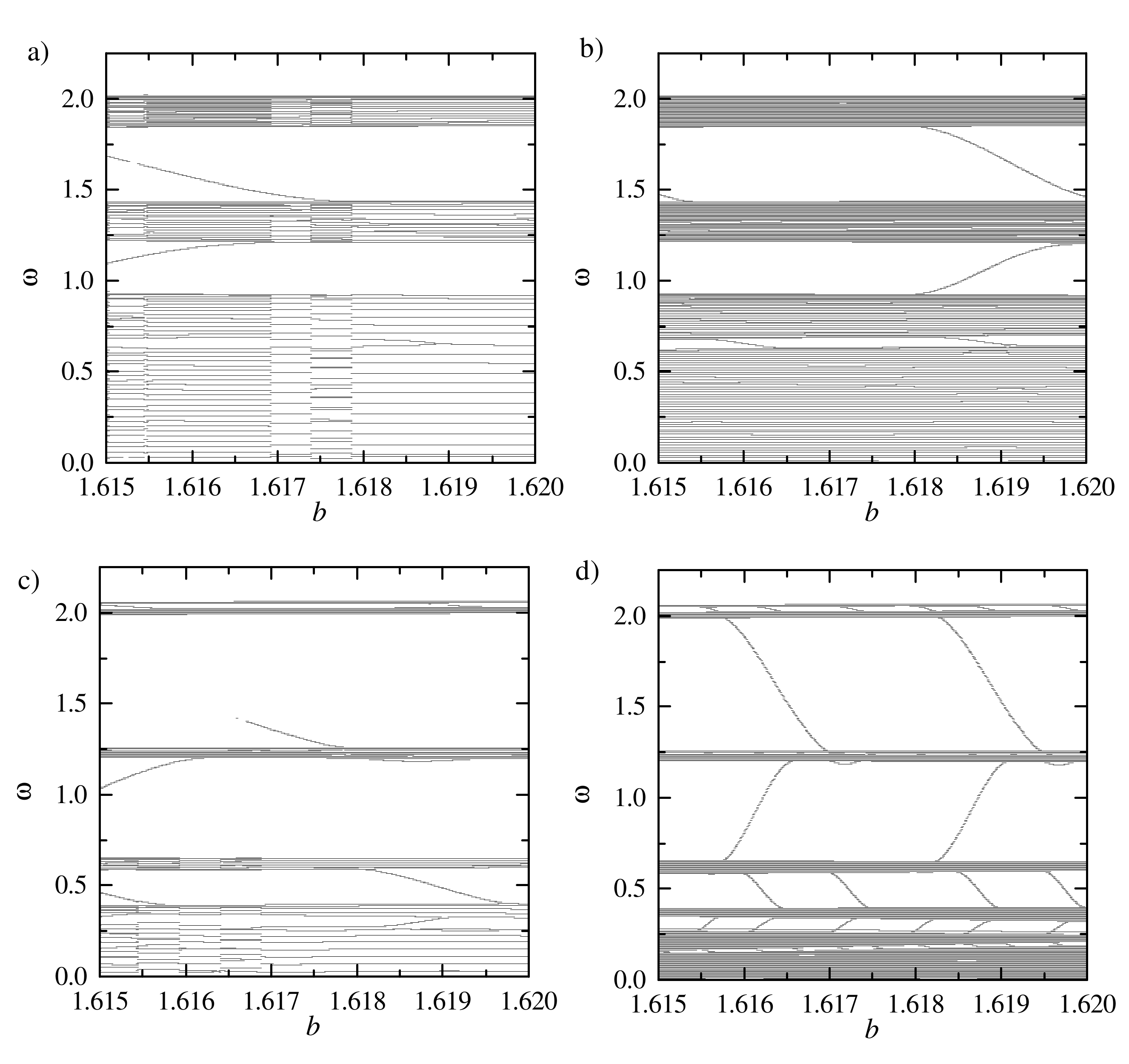}
\caption{Frequency spectrum around $b = 1.618$ for $\lambda = 0.5$ and $ \lambda = 1.0$, varying the number of sites on the network. We can see that the location of the states that cross the gaps is modified by the number of sites, and the bands are narrowed as we change $\lambda$}
\label{fig:ph-comp-N}
\end{figure}

 The phonon equivalent for the Hofstadter Butterfly has a strong influence on the number of sites and the $\lambda$ parameter. In Figure \ref{fig:ph-comp-N} we show the frequency modes around the dashed vertical line, in highlighted region of Figure  \ref{fig:ph-but-lbda-one} (red dashed vertical line in online color version). For $N = 100$ the bands cross only the smaller gaps, regardless of the two values of $\lambda$, while increasing the number of sites up to 206 the bands are narrowed, and these states cross all gaps. When $\lambda = 1.0$ and $N = 100$  from Figure \ref{fig:ph-comp-N} we verify the presence of forbidden bands for any approximation value for $b = \beta$,  within the analyzed range ($1.615$ up to $1.62$), similar to what occurs in the electronic case, where it appears only for the finite system and the origin of this effect is due to the conservation of the number of particles \cite{MADSEN2013}.

\begin{figure}[ht!]
\centering
\includegraphics[width=1.0\linewidth]{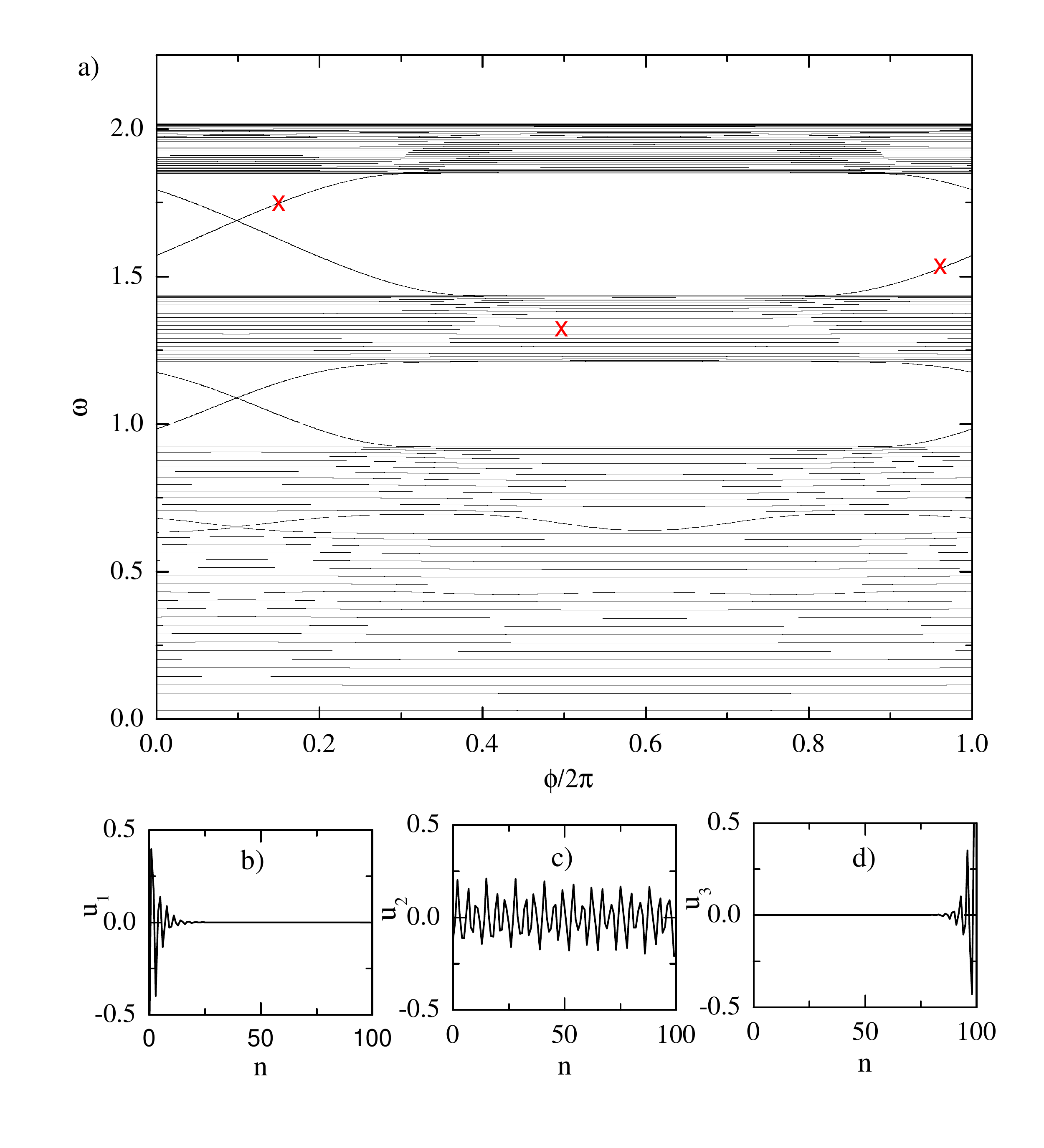}
\caption{Phonon frequency spectrum, as a function of the $\varphi$ phase. The parameters used were: $N = 100$; $\lambda = 0.5$. Below, we present the eigenvectors found at points marked with a  \textbf{\textcolor[rgb]{1.00,0.00,0.00}{$\times$}} on the spectrum.}
\label{fig:ph-egval-l-05-n100-u}
\end{figure}
\par The $\varphi$ phase also modifies the shape of the spectrum, as shown in Figure  \ref{fig:ph-egval-l-05-n100-u}, where we consider a lattice with $100$ atoms and $\lambda = 0.5$. We can notice that the frequencies are distributed in four separate intervals with larger gaps. Only a few modes cross the forbidden frequency gaps, with an almost sinusoidal dispersion. Below of the main painel, we can see the displacements $u_n$ (eq. 2) against the index $n$, calculated for three case, characterized by the (red) cross  in main painel on  the Figure \ref{fig:ph-egval-l-05-n100-u}, namelly $(u_ {1})$, $(u_ {2})$  and to the right $(u_3)$. We can se that $u_ {1}$ and $u_ {3}$ modes are strongly localized at the edges of the system, while the calculated mode for the center, labeled by $u_ {2}$, it is extended mode through all sites. The modes  $u_ {1}$ and $u_ {3}$ represent the topological states of phonons in this system. They are formed by oscillations at the edge of the quasiperiodic lattice for a set of well specified parameters $\phi $, $\lambda $ and $ N $ (For a review \cite{KRAUS2012}).
\par
\begin{figure}
\centering
\includegraphics[width=1.0\linewidth]{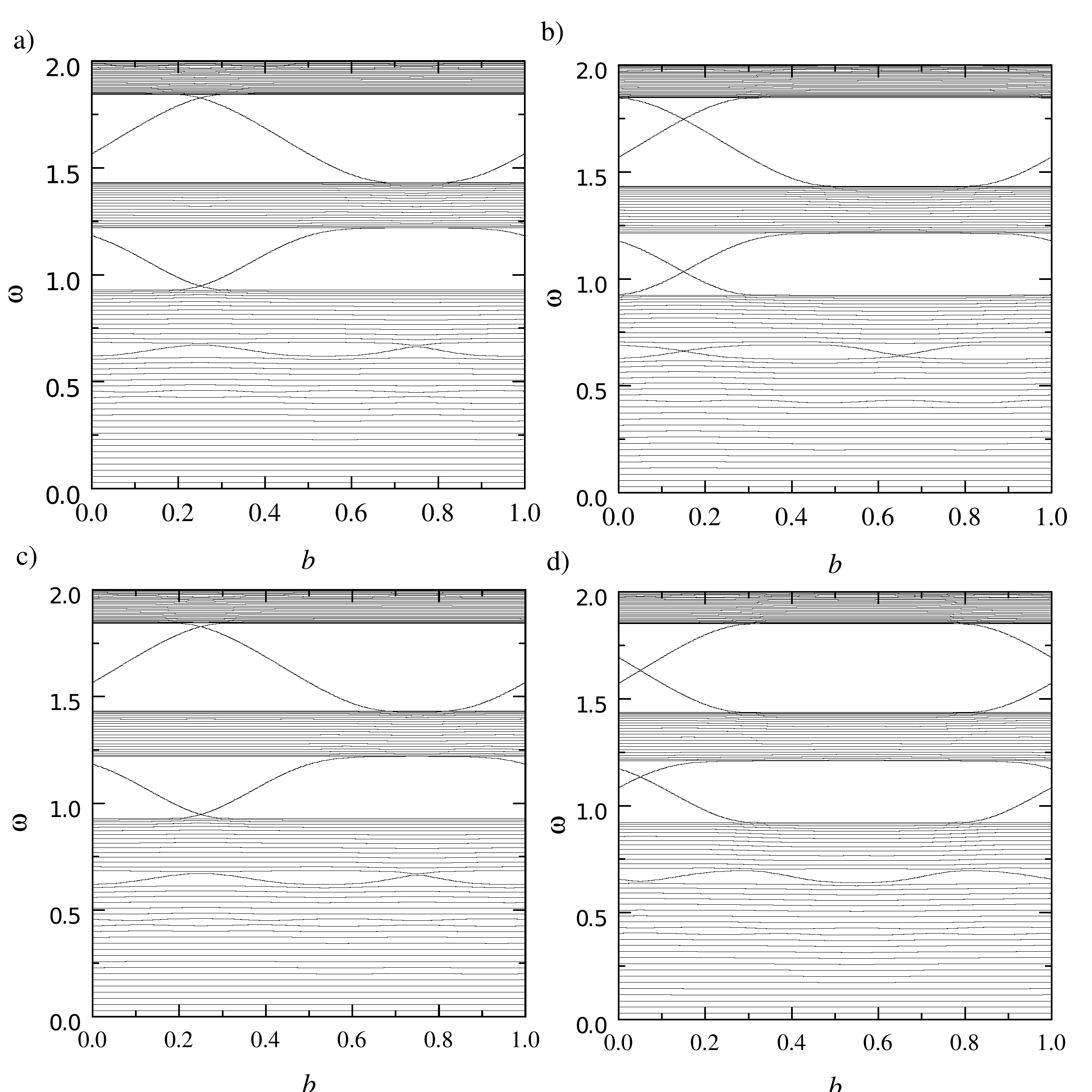}
\caption{Frequencies as functions of the $\varphi$ phase, obtained for changes in the value of $b$ in the third decimal place with $N = 100$ and $\lambda = 0.5$}
\label{fig:phonon-b-precision}
\end{figure}

 The numerical precision for the inverse of the frequency ($b$) in the force constant $K _{n}$ considerably alters the allowed eigenvalues in the spectrum as a function of the potential phase, as shown in Figure \ref{fig:phonon-b-precision}. In this way, we can see that the phonon states remain crossing the gaps, however, a translation occurs in the modes, and also a gentle deformation can be observed. All the eigenvalues were obtained from the equation \eqref{eq:matrix-fonon}, and they were calculated for four different $b$ approximations, for $\varphi$ values between $0$ and $2 \pi$. The border states that emerge in the larger gaps move, arising for different phase values.

\par
\begin{figure}[ht!]
\centering
\includegraphics[width=1.0\linewidth]{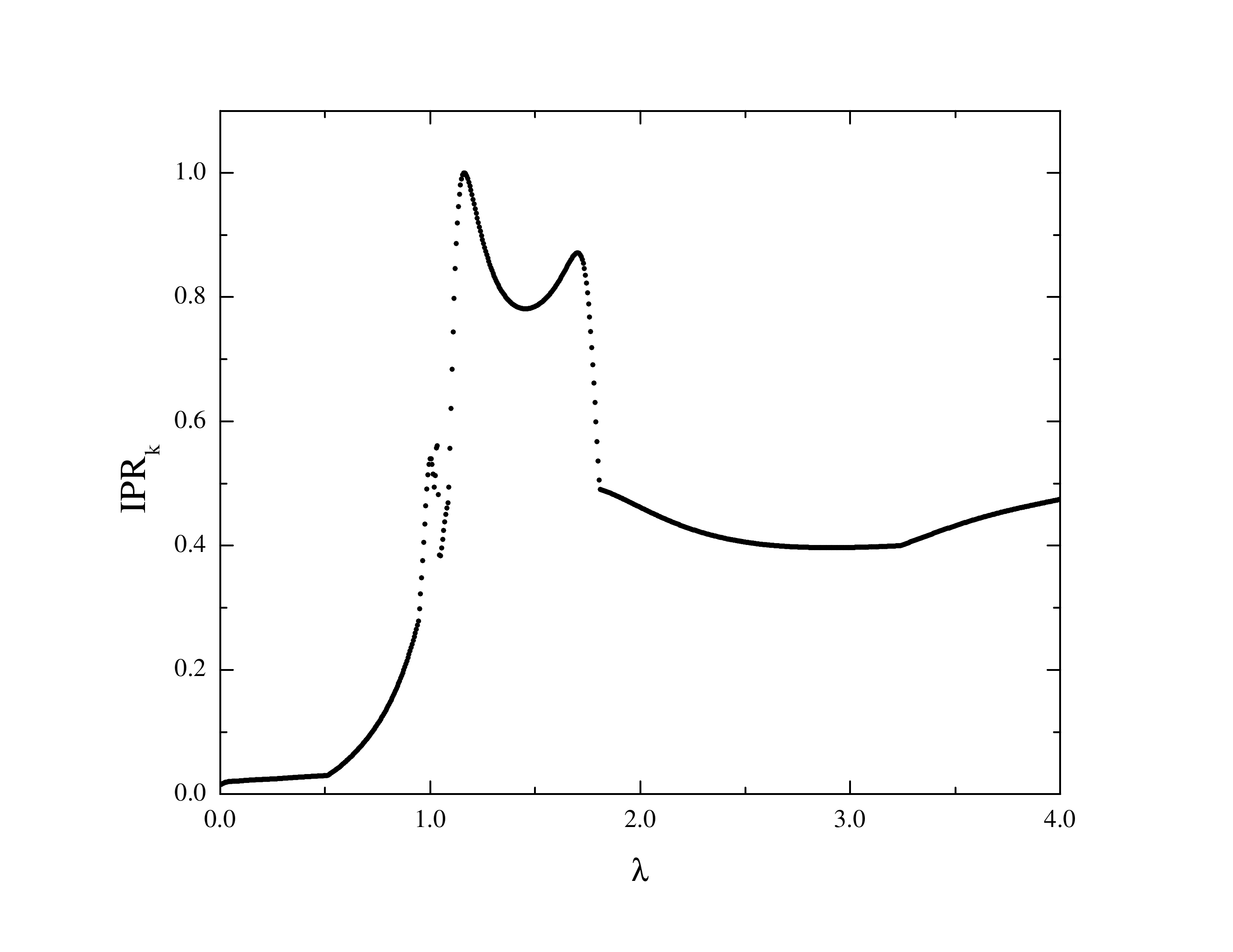}
\caption{Inverse of the participation rate for displacements with $N = 100$ and frequencies in the gap ($1.7$ up to $3.0$), showing a transition between displacements for values of $\lambda >1.0$}
\label{fig:ipr-phonon-n100}
\end{figure}

 The location of the frequency ranges within the upper gap depends heavily on the $\lambda$ parameter, which represents the amplitude of the cosine modulation in eq. (3). Clearly, from Fig. 2 we can see that there is two behavior in the energy (frequency) spectra, banded (where the bands are well defined) or unbanded (where the states are very narrow sets so that it is impossible to define it as a band) spectrum, depending on the parameter $\lambda$. It will play an important role in the referred phase transition. In order to study this
phase transition,  in Figure \ref{fig:ipr-phonon-n100}  we present the inverse of the participation rate for the frequency values located within the upper gap (frequency values greater than $1.7$).
We can see that, depending on the $\lambda$ parameter, the IPR can present a phase transition in the displacements $u_{n}$ of the lattice. For values smaller  than $ \lambda = 1.0$ the displacements are scattered across all sites representing extended states of the system, but as we increase the value of $\lambda$, the IPR shows an intense localization (high IPR), representing a transition in the system's oscillations,  from extended displacements to localized oscillations.

\begin{figure}[ht!]
\centering
\includegraphics[width=1.0\linewidth]{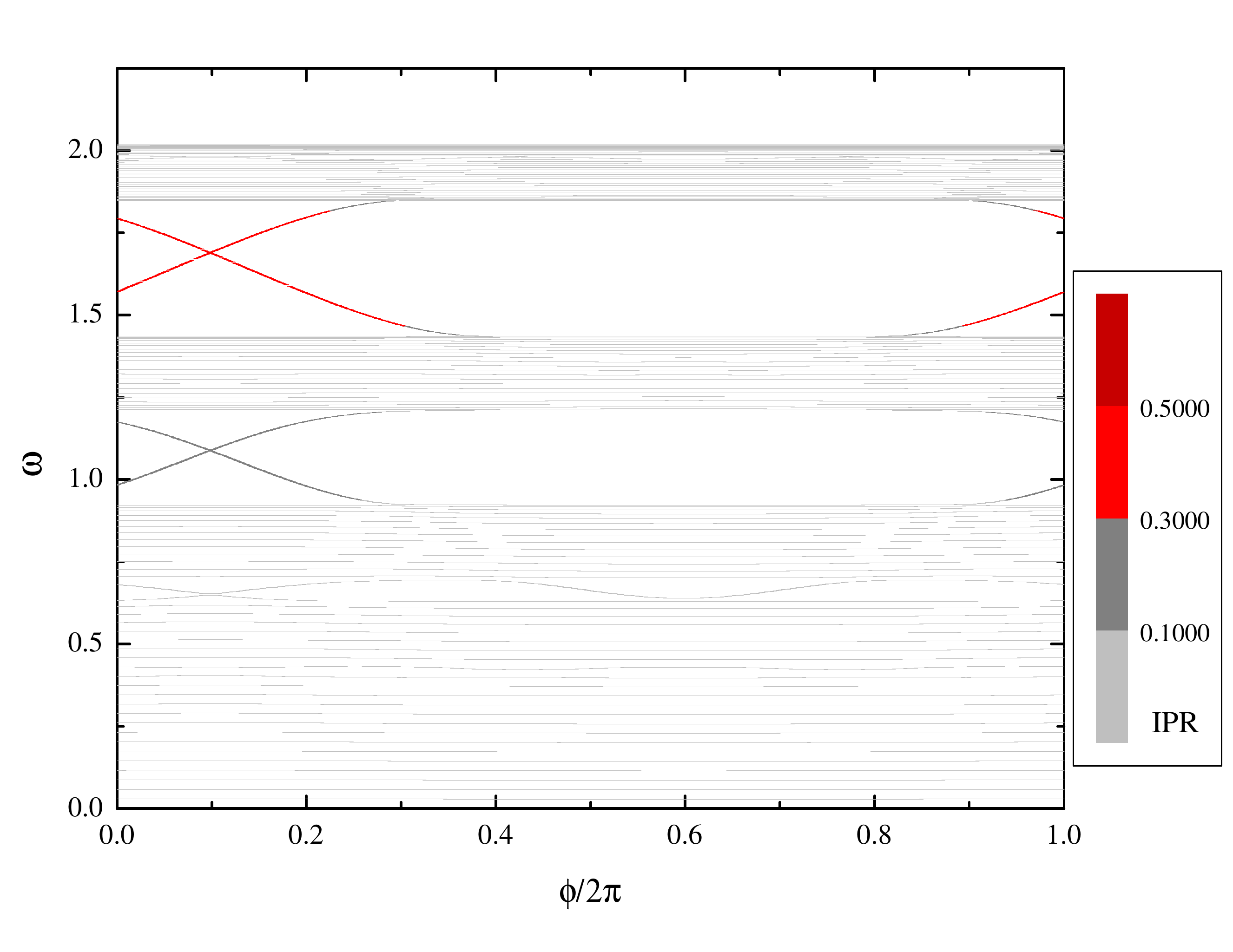}
\caption{Frequency spectrum as a function of the $\varphi$ phase of the potential with IPR in color scale for $ \lambda = 0.5$ in a lattice with 100 atoms.}
\label{fig:ph-w-X-phi-IPR}
\end{figure}

\begin{figure}[ht!]
\centering
\includegraphics[width=1.0\linewidth]{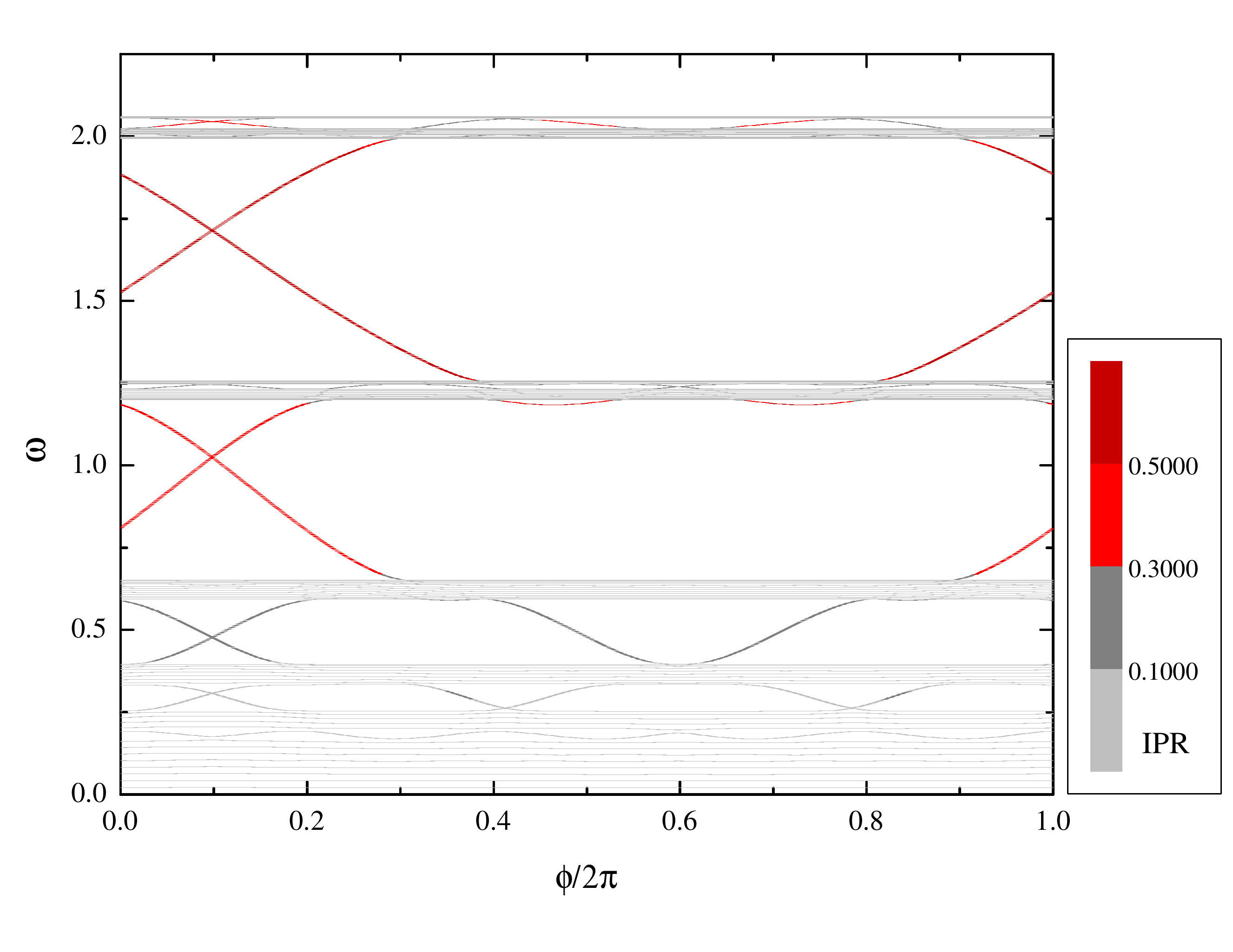}
\caption{Frequency spectrum as a function of the $\varphi$ phase of the potential, with IPR in color scale for $ \lambda = 1.0$, in a lattice with 100 atoms.}
\label{fig:ph-w_phi-lbda09}
\end{figure}

The location of the displacements in the phonon spectrum is also modified, as a function of the $\varphi$ phase, as we can see in Figure  \ref{fig:ph-w-X-phi-IPR},  where we added the gray (color, in online version) scale for a system with $100$ sites and $\lambda = 0.5$. The states that cross the second largest gap present a more intense localization between the gaps in this figure (higher IPR, dark less color), while the remainder is fully extended (more gray, or red in color version, i.e., lower IPR; see Fig. \ref{fig:ph-egval-l-05-n100-u}).
When we vary the $\lambda$ parameter up to the critical value ($\lambda = 1.0 $), the set of extended bands are narrowed and the states that cross the larger gaps are more localized, as seen in Figure \ref{fig:ph-w_phi-lbda09}.
The frequency range where the gaps exist are very close to that ones in Figure \ref{fig:ph-w-X-phi-IPR}, mainly in the upper gap, not altering, therefore, the frequency range for emergence of the topological states.

\begin{figure}[ht!]
\centering
\includegraphics[width=1.0\linewidth]{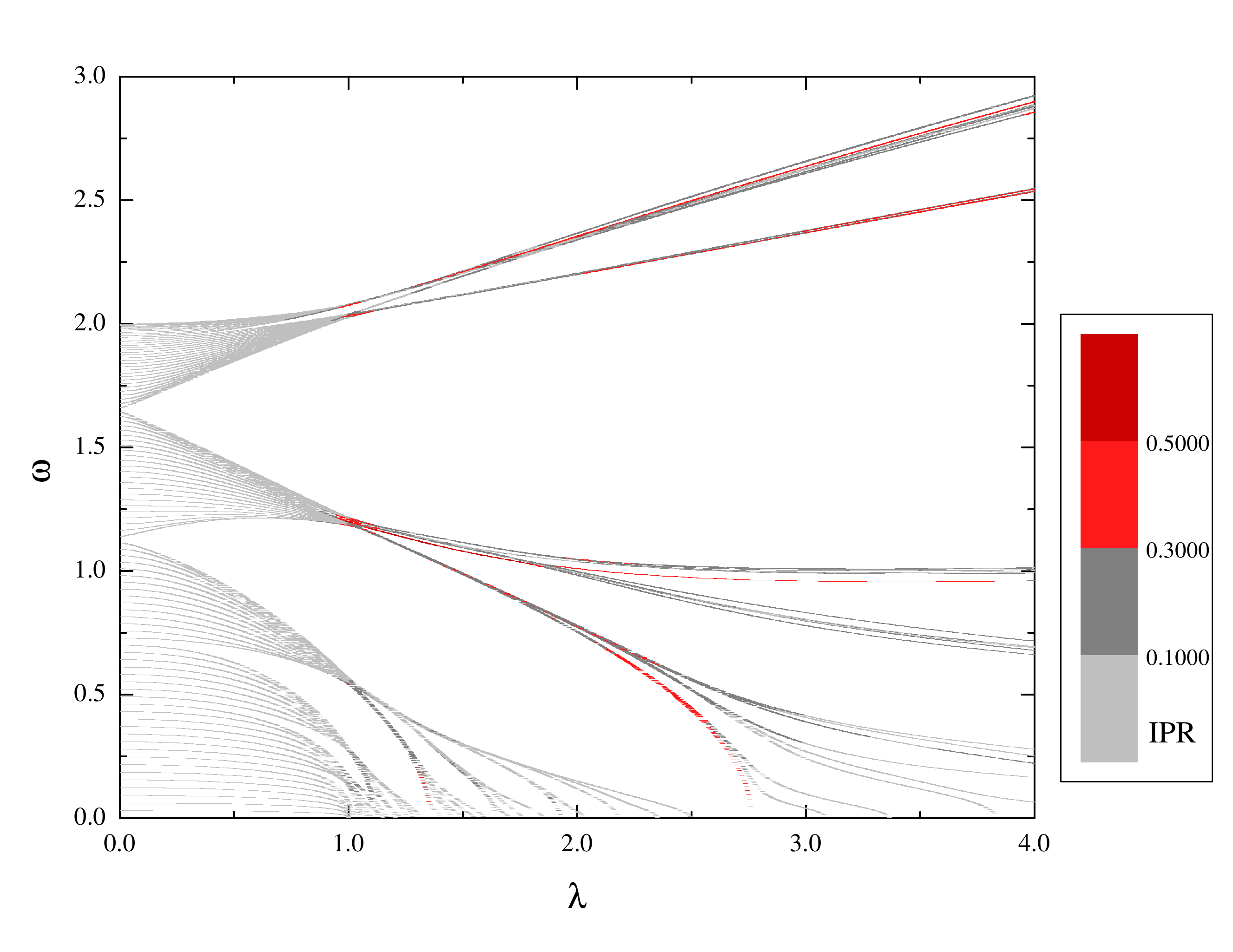}
\caption{Frequency dispersion as function of the amplitude parameter $\lambda$, with $100$ atoms in the lattice, considering the initial phase $\varphi = 0$. We can see that for values greater than $\lambda=1.0$ the spectrum presents many narrow bands spread with hight band gaps, and for,  less than $1.0$, we have more larger and well orgnized bands, characterizing $ \lambda=1.0$, as a critical value for a phase transition. }
	\label{fig:spctrxlbda}
\end{figure}

In Figure \ref{fig:spctrxlbda} we present the frequency dispersion as a function of the amplitude of the potential, incommensurate with the IPR, in color scale. We can see that for $ \lambda<1.0$ the frequencies present itself as bands separated by larger gaps, while for $\lambda=1.0$ the spectrum comes down to vibration modes at specific and well-defined frequencies. The $\lambda$ parameter allows one to control the distribution of bands in the spectrum, representing a multifractal when they are close enough ($ \lambda = 1.0 $) and deforming the spectrum for higher values of $\lambda$, organizing itself in modes of vibration with wide gaps.

\section{Conclusions}
The system studied here consists of an adaptation of the Hamiltonian of Aubry-André to deal with the elementary vibrations of the unidimensional quasicrystalline lattice. The equation of the eigenvalues for our case consists of a coupled system, with the elementary oscillations in each site  interacting by a force constant given in the coupling with the neighboring atoms. We have used the numerical diagonalization method to find the allowed frequencies (computational package of GSL). We have found that the frequency spectrum is modified according to the $\lambda$ interaction parameter, presenting the phonon equivalent for the Hofstadter Butterfly's for $\lambda= 1.0 $, keeping the symmetry and gaps close to the one obtained in the electronic case. Also, we have verified that the number of atoms in the network influences the number of gaps and edges modes that cross these gaps. For the very precise value of the $b$ parameter (around $b = 1.618$), and choosing a given $\lambda$, it is possible to control these frequency bands.

The interaction with this quasicrystalline potential can be modulated by the $\varphi$ phase, causing the atoms on the edge of the system to vibrate with one modulated frequency lying between the gaps, characterizing an edge mode. It is well known that topological phases are characterized by edge states confined near the boundaries of the system, whose the modes are lying in a bulk energy (or frequency) gap \cite{2015Bahri}, and it is unaffected by disorder or deforming, for example. Therefore, we can infer that  due to the analogy with the photonic case studied by Kraus et. al \cite{KRAUS2012}, we have a phononic topological phase exhibiting the so-called ``topological states".

Indeed, the topological properties of the 1D quasicrystals can emerge in two ways \cite{KRAUS2012}: through the existence of quantum phase transitions when we have a continuously deforming between two topologically distinct quasicrystals or by the appearance of robust boundary states which traverse the bulk gaps as a function of some  controllable parameter (which in this case is the initial phase $\phi$). Specifically, we have considered the second way in a 1D phononic quasicrystal, on order to show that is possible to have localized boundary states (edges states), which manifest the same topological properties of a 2D photonic quasicrystal \cite{PhysRevX.6.011016}. Therefore, it is possible to define an equivalent Chern number for our case and classify topologically the states of the system studied here\cite{KRAUS2012,Deymier2016}. We will consider this in furthers works. 

On the other hand, studying the individual displacements $u_{n}$, for a given frequency it is possible to characterize an equivalent metal-insulator phase transition. Therefore, as we can see studying the IPR (Fig. 7), this phase transition can be characterized by a critical value of $\lambda= 1.0$.

\begin{acknowledgments}

We would like to thank CNPq (Conselho Nacional de Desenvolvimento Cient\'{\i}fico e tecnol\'{o}gico) for the partial financing. This study was financed in part by the Coordenao de Aperfeioamento de Pessoal de Nível Superior - Brasil (CAPES) - Finance Code 88881.172293/2018-01.

\end{acknowledgments}

\bibliography{bibliograph}

\end{document}